\documentclass[12pt]{article}

\usepackage{natbib}
\usepackage{graphicx}
\usepackage{amsmath}
\usepackage{url}
\usepackage{bbm}

\title{Future Proofing a Building Design Using History Matching Inspired Level-Set Techniques}
\author{Evan Baker \thanks{Primary and corresponding author: Department of Mathematics, University of Exeter; \texttt{e.baker@exeter.ac.uk}; Laver Building, North Park Road, Exeter, EX4 4QF},
 Peter Challenor \thanks{Department of Mathematics, University of Exeter, \tt P.G.Challenor@exeter.ac.uk},
  and Matt Eames \thanks{Department of Engineering, University of Exeter \tt M.E.Eames@exeter.ac.uk}}
\date{}

\begin{document}

\maketitle

\begin{abstract}
How can one design a building that will be sufficiently protected against overheating and sufficiently energy efficient, whilst considering the expected increases in temperature due to climate change? We successfully manage to address this question - greatly reducing a large set of initial candidate building designs down to a small set of acceptable buildings.
We do this using a complex computer model, statistical models of said computer model (emulators), and a modification to the history matching calibration technique. This modification tackles the problem of level-set estimation (rather than calibration), where the goal is to find input settings which lead to the simulated output being below some threshold. The entire procedure allows us to present a practitioner with a set of acceptable building designs, with the final design chosen based on other requirements (subjective or otherwise).

\end{abstract}


\section{Introduction}
\label{sec:intro}

Our goal is to find modifications to an existing building design which will provide satisfactory overheating risks \emph{and} energy demands; even after the expected increase in temperature by the end of the 21st century. We aim particularly for this notion of `satisfactory' rather than `optimal', because in practice there are often additional criteria when it comes to building design (such as its appearance). Additionally it is far easier, and more sensible, for regulation to require a specific threshold standard than it is to require some relative notion of `most-improved'.

To aid in this goal, we use EnergyPlus \citep{Crawley2000}, which is a numerical model (a simulator) for simulating many properties of a given building, such as its total annual energy usage, or its hourly temperature. Because EnergyPlus takes time to perform a single simulation (the specific amount depends on the complexity of the building) and our version of EnergyPlus is stochastic, we use statistical models (emulators) to aid in our analysis. Statistically modelling complex computer models is a well researched idea \citep{Sacks1989, Kennedy2001, Oakley2002}, with perhaps the most widespread emulator being the Gaussian process emulator, which interpolates previously obtained simulations and provides uncertainty estimates for these interpolations \citep[see][for a tutorial]{OHagan2006}. 

Emulators can have many different uses. Perhaps the two most explored applications are prediction (quickly predicting what the simulator output is for new inputs), and calibration (finding the `true' input values using observational data). For calibration, standard Bayesian inference provides one such solution \citep{Kennedy2001}, but it is not without its flaws \citep{Brynjarsdottir2014}. An alternative method, history matching \citep{craig1997pressure, Vernon2014, Andrianakis2015}, also exists, providing straightforward general implementation, easy utilisation even when the input and output dimension is high, a robustness to low simulation budgets, and the ability to identify when no such `real' input exists. 

With our goal, we are interested in using emulators to perform level-set estimation. The `level-set' typically refers to the set of inputs where the output is exactly some threshold, but we use the term more liberally to refer to when the output is less than some threshold (finding the former implicitly reveals the latter, and vice-versa). History matching techniques have already been extended to the problem of optimisation \citep{Lawson2016}, and so it is a natural step to extend them to level-set estimation as well. 

Our methodology is intrinsically very accessible, and easily applied to new problems. This makes it well-suited as a framework for engineers intending to `future-proof' buildings. We think this case study provides interesting problems, and the field of building performance simulation is crying out for greater attention from statisticians in general. Further details about the motivating problem, and the specific building we use as our example is provided in Section \ref{sec:simul}. Section \ref{sec:emul} outlines the emulators used, and the various nuances provided by the problem. Section \ref{sec:HMLSE} provides an explanation of our history matching inspired method for level-set estimation, and Section \ref{sec:results} applies the method to the discussed building model application. Section \ref{sec:conc} then provides some concluding remarks.

\section{Building Model}
\label{sec:simul}

\begin{figure}[!htb]
\includegraphics[width=\textwidth]{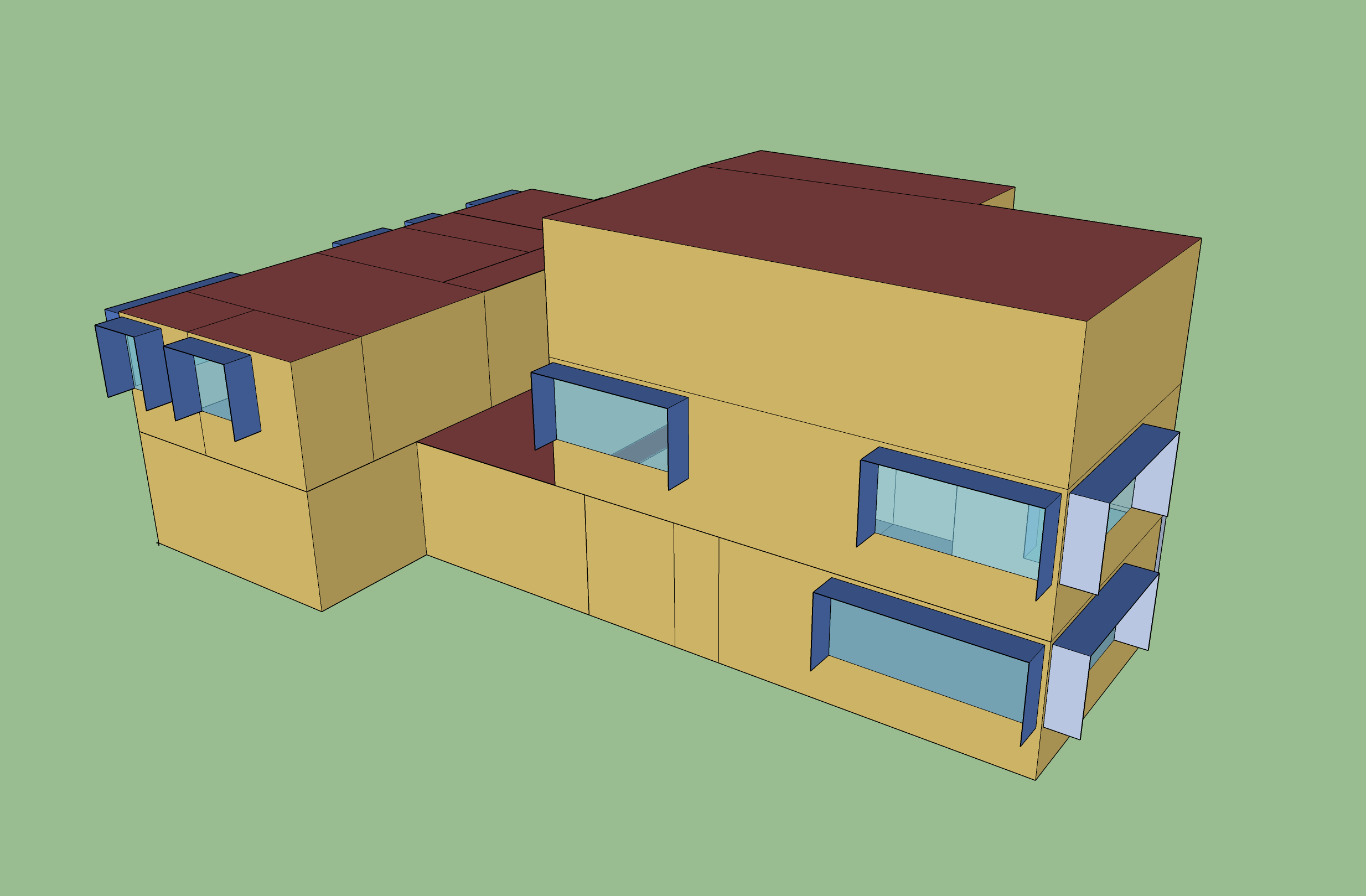}
\caption{The geometry of the modelled building.}
\label{fig:building}
\end{figure}

EnergyPlus requires the shape and design of the building in question to be provided. For the purposes of this article, we use a specific building design (an image of this design is given in Figure \ref{fig:building}), but the process outlined is not specific to this example building. 

Our building is fitted with an ``ideal loads" heating system, and no air conditioning. No air conditioning may seem like an odd setup, given the intended objective, but this choice is more representative of the UK's building stock, where it has been reported that only $0.5\%$ of residential buildings have air conditioning \citep{BBC2013}. Air conditioning could also have been included, with a user-input capacity, without any major methodological modifications.

To improve this building, we can edit several properties of the building design (for example, the thickness of insulation). In this work, we consider editing the following inputs: the wall insulation thickness (considering values from 0m to 0.5m), the roof insulation thickness (from 0m to 0.5m), the ground insulation (from 0m to 0.1m), the size of the windows (from $20\%$ to $100\%$ of the wall size, ensuring at least a 0.1m windowless border), the length of window overhangs (from $0\%$ to $100\%$ the height of the windows), the amount the windows can be opened by occupants (from $0\%$ to $100\%$ of the windows size), the emissivity of the roof (from 0.4 to 1), and whether or not the windows are double or triple glazed. From now on, these are referred to as $x_1, \dots, x_8$ and their input ranges are rescaled to be between 0 and 1. Other inputs could have been considered (for example, the air conditioning capacity); the specific choices in practice are down to what construction options are available. 

EnergyPlus also requires the outside weather input. Standard practice takes this weather as a fixed, known thing \citep{Eames2015, Eames2016b}, and so EnergyPlus is usually deterministic. In our work, we opt instead to take the weather as random: each time EnergyPlus is run, a new sample of weather is drawn from a random weather generator, more accurately representing the chaotic and uncertain nature of weather. From here on we refer to this whole procedure as EnergyPlus, and so our simulator is now stochastic. Our interest lies in `future-proofing' the building, i.e. making sure it performs well towards the end of the 21st century. With this in mind, our specific choice of weather generator is the UKCP09 weather generator, which can output possible samples of weather for the year 2080, which can then be input into EnergyPlus \citep{Eames2011}. In this way, the choice of weather generator directly affects the analysis made - other weather generators could have been chosen (or even fixed weather could be used) were the goals different.

After providing the building design and the weather, EnergyPlus can simulate the yearly heating energy usage and the indoor temperature throughout the year. The indoor temperature can then be used to classify whether or not the building overheated during the simulation. This classification of overheating is defined by the Chartered Institution of Building Services Engineers (CIBSE,  \cite{cibse201352}), and a short description of this classification is provided in the appendix. Other classifications or metrics for overheating could have been considered instead.

Because our modified EnergyPlus is stochastic, any single simulation is not particularly informative. A building may not overheat, or have low heating costs, only because the weather was favourable in that simulation. We instead want to reduce the \emph{probability} of a given building design overheating, and the \emph{average} heating energy usage of a given building design. In this work, we semi-arbitrarily decide to aim for a less than $1\%$ probability of overheating. This represents a `sufficiently small' value, but a different value could have been chosen. For the average heating energy usage, we aim for less than $15 \mathrm{kWh}/\mathrm{m}^2$, which is the requirement set by the passivhaus standard \citep{Schnieders2006}, but a different threshold could also have been chosen.

To summarise, our goal is to find the set $\mathcal{L}$, such that:

\begin{equation}
\label{eq:levelset}
\mathcal{L} = \{(x_1, \dots, x_8) : P\left( y_{oh}(x_1, \dots, x_8) \right) <0.01, E\left( y_{eu}(x_1, \dots, x_8) \right) < 15\}
\end{equation} 

where $(x_1, \dots, x_8)$ are the different design choices, $y_{oh}$ is the binary output that classifies whether the building overheats and $y_{eu}$ is the building's energy usage. 

\section{Emulators}
\label{sec:emul}

Constructing emulators for our two outputs is essential. Not only does EnergyPlus take time to run (the building above takes roughly 1 min for one simulation), which can be alleviated by emulators, but the outputs of interest (overheating \emph{probability} and \emph{average} energy usage) are not directly provided by EnergyPlus. This restriction demands some degree of statistical modelling, and so what follows is a description of how we modelled these outputs.

The existence of the binary input variable $x_8$ is a mild complication. This is non-standard in the emulation community, with a standard Gaussian process formulation requiring all inputs be continuous. The window glazing variable was selected partially to show that binary input variables can still be included, as binary variables are likely to be common as potentially adjustable attributes in a building design. In this work, we use the mechanism outlined in \citep{Qian2007} that allows non-continuous variables to be included in the covariance structure of a Gaussian process. A simpler (but less efficient) alternative would have been to fit independent emulators for each of the different window glazing options.

We first model the overheating risk using a Gaussian process classifier \citep{Rasmussen2006}. This models overheating as a Bernoulli variable with a latent Gaussian process for the logit probability of overheating ($logit(p)$). Taking the continuous inputs as $\textbf{x}_c = {x_1, \dots, x_7}$, and the binary input(s) as $\textbf{x}_b = {x_8}$, we have the following overheating risk emulator:

\begin{equation}
\label{eq:classifier}
\begin{gathered}
y_{oh}(\textbf{x}_c, \textbf{x}_b) \sim Bernoulli(p(\textbf{x}_c, \textbf{x}_b))\\\\
\mathrm{logit}(p(\textbf{x}_c, \textbf{x}_b)) \sim GP(m_{oh}(\textbf{x}_c, \textbf{x}_b), K_{oh}(\textbf{x}_c, \textbf{x}_b, \textbf{x}_c', \textbf{x}_b'))
\end{gathered}
\end{equation}

The mean function $m_{oh}(\textbf{x}_c, \textbf{x}_b)$ models the overall trend, and can be used to provide parametric prior beliefs about the input-output relationship. The covariance function $K_{oh}(\textbf{x}_c, \textbf{x}_b, \textbf{x}_c', \textbf{x}_b')$ provides a correlation structure, allowing more nuanced local details to be captured. We use a zero-mean function, letting the covariance function do all of the work. The covariance function is taken as:

\begin{equation}
\label{eq:correl}
K_{oh}(\textbf{x}_c, \textbf{x}_b, \textbf{x}_c', \textbf{x}_b')) = \alpha_{oh}^2 exp\left( - \sum_{x \in \textbf{x}_c} \frac{(x_i - x_i')^2}{l^2_{oh_i}} - \sum_{x \in \textbf{x}_b} \phi_{oh_i} \mathbbm{1}(x_i \neq x_i')\right)
\end{equation}

$\alpha_{oh}^2$ is the overall variance of the process, and represents the overall epistemic uncertainty. The lengthscales $l_{oh_i}$ control the correlation between two points according to how far apart they are with regard to the continuous inputs (with each input dimension $i$ having a corresponding lengthscale). Each correlation parameter $\phi_{oh_i}$ controls the correlation between two points according to whether the $i^{th}$ binary input differs ($\mathbbm{1}$ is the indicator function). The left summation disappears if the two inputs have the same values for the continuous variables, and the right summation disappears if the two inputs have the same values for the binary variables. In our case, we only have one binary variable, so the right summation can be replaced with a single term, but the full summation is provided here for generality.

For hyperparameter priors, the overall variance is given a $Half-Normal(0,1)$ prior, the lengthscales are given $Inverse-Gamma(5,5)$ priors and the binary correlation parameter is also given a $Half-Normal(0,1)$ prior.

Fitting this model is done via variational inference using GPflow \citep{GPflow2017}, providing accurate uncertainty distributions for the latent probability estimates.

The second emulator is the one for energy usage. We model this with a heteroscedastic Gaussian process \citep{Kersting2007, Boukouvalas2010, Binois2018}, which models the mean as a Gaussian process and the (log) variance as a Gaussian process. Using the same notation as before we have the following energy usage emulator:  

\begin{equation}
\label{eq:emulator}
\begin{gathered}
y_{eu}(\textbf{x}_c, \textbf{x}_b)|\delta^2(\textbf{x}_c, \textbf{x}_b) \sim GP(m_{eu}(\textbf{x}_c, \textbf{x}_b), K_{eu}(\textbf{x}_c, \textbf{x}_b, \textbf{x}_c', \textbf{x}_b') + \delta^2(\textbf{x}_c, \textbf{x}_b)) \\\\
log(\delta^2(\textbf{x}_c, \textbf{x}_b)) \sim GP(m_{\delta}(\textbf{x}_c, \textbf{x}_b), K_{\delta}(\textbf{x}_c, \textbf{x}_b, \textbf{x}_c', \textbf{x}_b'))
\end{gathered}
\end{equation}

Where $\delta^2$ represents the intrinsic variability of the energy usage output. The mean functions ($m_{eu}(\textbf{x}_c, \textbf{x}_b)$ and $m_{\delta}(\textbf{x}_c, \textbf{x}_b)$) and covariance functions ($K_{eu}(\textbf{x}_c, \textbf{x}_b, \textbf{x}_c', \textbf{x}_b')$ and $K_{\delta}(\textbf{x}_c, \textbf{x}_b, \textbf{x}_c', \textbf{x}_b')$) are given the same structure and priors as the overheating emulator. This model is fit via maximum a posteriori estimation using Stan \citep{Team2015}, but the Gaussian process structure ensures that a full probability distribution is still obtained for the mean (which is our quantity of interest).  If it weren't for our non-standard covariance function, the hetGP R package would have been suitable here \citep{hetGP}.

Together, these two emulators provide a way of predicting what the overheating risk and the average energy usage are for any values of $(\textbf{x}_c, \textbf{x}_b)$. These predictions will have an uncertainty distribution around them which can easily be obtained from the Gaussian process predictive equations. The accuracy and precision of these predictions will depend on the total number of simulations used to fit these models.

The next section will outline the history matching inspired level-set estimation methodology - detailing how one can better estimate the level-set of a simulator. The section after that will then apply said methodology to the above emulators - finding suitable buildings with regard to overheating risk and average energy usage. One key benefit of the proposed methodology, which is worth noting now, is that it is easily generalisable to many types of emulator - as long as a mean and variance of a prediction can be provided, then the proposed methodology can be directly used.

\section{History Matching Level-Set Estimation}
\label{sec:HMLSE}

We adapt the history matching procedure to apply to level-set estimation, rather than calibration. More information about standard history matching can be found in \cite{craig1997pressure, Vernon2014}; and \cite{Andrianakis2015}. The general idea here is to build initial emulators for our quantities of interest, and use these to rule-out buildings which almost certainly do not meet the desired criteria. The emulators can then be improved by obtaining more simulations for building designs which have not yet been ruled-out, providing a better estimate for what the level-set is. This process can be repeated, each time improving the results. Each set of simulations and analysis is typically referred to as a `wave'.

Taking $y$ as the quantity of interest (overheating risk or average energy usage), $\textbf{x}$ a set of inputs ($(x_1, \dots, x_8)$) and $T$ the threshold (0.01 or 15), we define the the `implausibility' as follows:

\begin{equation}
\label{eq:Impl}
I(\textbf{x}) = \frac{E(y(\textbf{x})) - T}{\sqrt{V(y(\textbf{x}))}}
\end{equation}

Where $E(y(\textbf{x}))$ is the expectation of the emulator, and $V(y(\textbf{x}))$ is its variance. This implausibility can be positive or negative, which is a key difference to standard history matching. Large, positive, values of $I(\textbf{x})$ suggest the input setting is not in the level-set, as the expectation is much larger than the threshold $T$. Large, negative, values suggest that it is in the level-set, as the expectation is much smaller than $T$. 

A value of 3 or greater for $I(\textbf{x})$ is taken as the threshold for a value being `implausible'. Three is the value often used in standard history matching, based on the Pukelsheim's three sigma rule \citep{Pukelsheim1994}. Any value of $\textbf{x}$ with $I(\textbf{x}) > 3$ is ruled-out, and no longer needs to be considered. The set of $\textbf{x}$ values that are not ruled out yet is often referred to as the `NROY' space (the Not Ruled-Out Yet space). Similarly, we `rule-in' any value of $\textbf{x}$ where $I(\textbf{x}) < -3$; these are so likely part of the level-set that they are not worth wasting further simulations on, and thus also no longer need to be considered (but do need to be remembered). The set of $\textbf{x}$ values which are not ruled-in yet will hereon be referred to as the `NRIY' space (the Not Ruled-In Yet space), which is not present in standard history matching.

\begin{figure}[!htb]
\includegraphics[width=\textwidth]{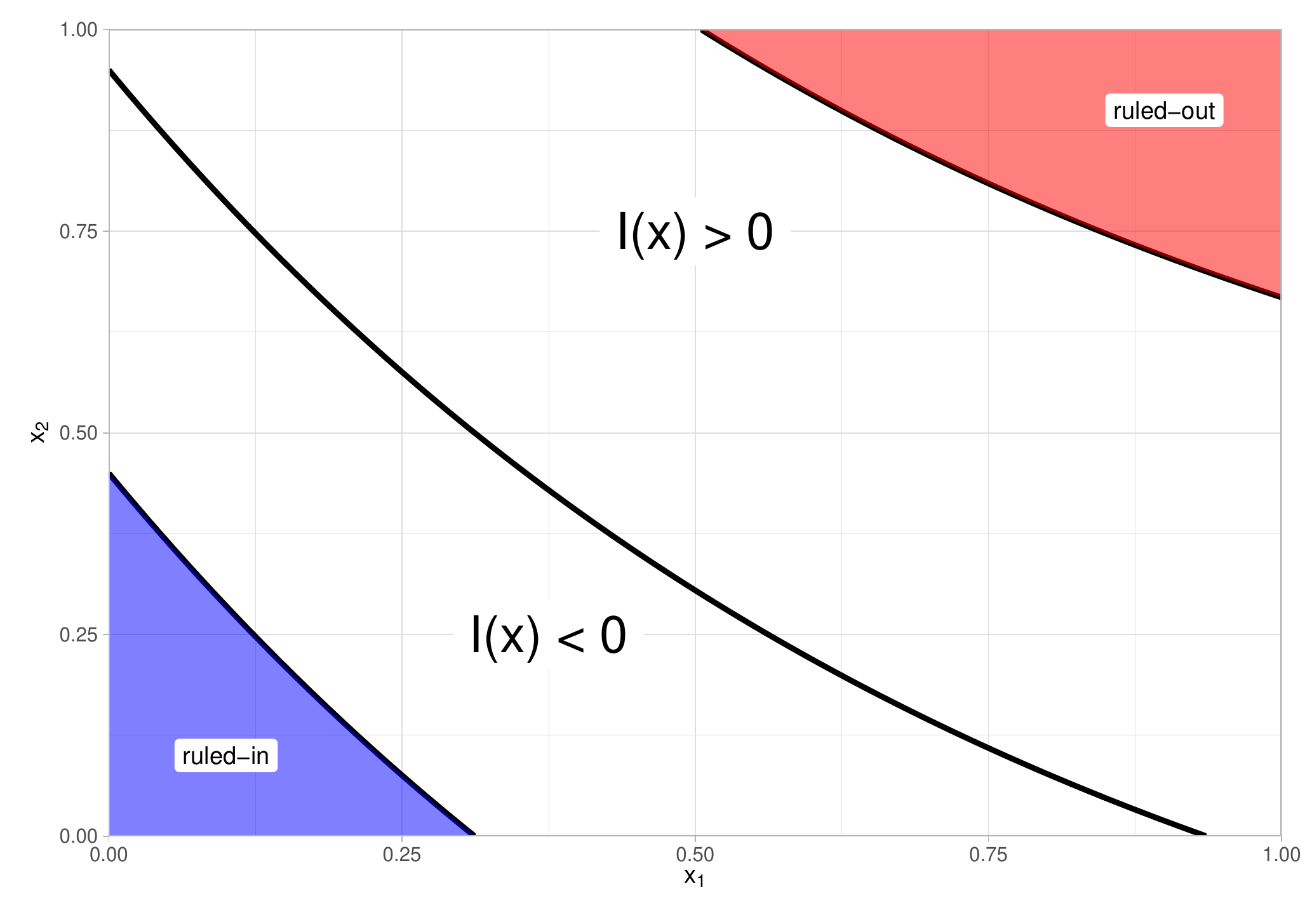}
\caption{An illustration of the 4 regions that emerge from the history matching level-set estimation technique. The non-blue regions represent the NRIY space, and the non-red regions represent the NROY space. As such, the white regions represent the space where inputs are both NROY and NRIY.}
\label{fig:NROYNRIY}
\end{figure}

For clarification, consider the image in Figure \ref{fig:NROYNRIY}.
This illustration demonstrates how 4 distinct regions of space emerge from using the implausibility metric from Equation \ref{eq:Impl}. The central line, going from the top left corner to the bottom right corner, represents the set of inputs where the output exactly equals $T$. The red, top right, region represents the set of inputs where the output is much larger than $T$; these are therefore almost certainly not in the level-set, and thus are ruled-out. The blue, bottom left, region represents the set of inputs where the output is much smaller than $T$; these are therefore almost certainly in the level-set, to the extent that they become uninteresting, and thus are ruled-in. The uncoloured middle regions are the regions of greater interest. The upper uncoloured region, NROY, represents the set of inputs where the implausibility is greater than 0, and thus are not believed to be in the level-set; but the implausibility is not large enough to know for certain. On the other hand, the lower uncoloured region, NRIY, represents the set of inputs where the implausibility is smaller than 0, and thus are believed to be in the level-set; but the implausibility is not small enough to know this for certain either.

With this, we then have a set of $\textbf{x}$ values which are candidates for future simulations (any values where $-3 < I(\textbf{x}) < 3$). Running simulations for some of these NROY / NRIY values and refitting the emulator will improve the emulator in this space. This space is the region closest to the boundary, and additional simulations here should help to distinguish which side of the boundary specific buildings are. This process can then be repeated several times, until the space of NROY / NRIY is acceptably small, it does not appear to change, or the simulation budget is exhausted.

If at any point, no choices of $\textbf{x}$ are NROY (i.e. all values of $I(\textbf{x})$ are greater than three), then this implies that no values of $\textbf{x}$ are in the level-set.

If more than one output is being emulated (as in our problem, where we have two quantities of interest), one can take the overall implausibility to be the maximum of the individual implausibilities - if it is implausible that a specific building meets one of the individual criteria, then that building is considered implausible overall.

In the final wave, when a final decision must be made (or a set of final candidate values must be presented to a practitioner), it is not reasonable to allow the choice of any `non-implausible' values if they still have fairly large implausibility, but not quite as large as three. Such buildings are still predicted by the emulators to fail the criteria, but not with enough certainty to rule them out. Therefore, in the final wave, we can impose stricter requirements, such as any input values where the implausibility is less than 0 (which we call the `tenable' set), or inputs where the implausibility is less than -1. A more desirable choice would be to only consider values with an implausibility less than - 3 (i.e. those ruled-in), with our simulation budget however we find this to be too strong a requirement as almost none of our candidate buildings end up ruled-in.

This methodology is exceedingly easy to implement, and is conceptually understandable - we rule out values that are obviously not in the level-set, rule in those that are obviously in the level-set, and all others can be investigated further.

In the next section, we shall apply this methodology to the two building criteria described previously.

\section{Results}
\label{sec:results}

We start by considering (but not simulating) a set of 1000000 candidate buildings (these are chosen by constructing two random latin hypercubes of size 500000 \citep{mckay1979comparison}, one for each value of the binary input variable). Our goal is to reduce this huge number of candidate buildings, to a more manageable subset of `future-proofed' buildings. For clarity, the different inputs and their initial considered ranges are presented again in Table \ref{tab:inputs}

\begin{table}[!htb]
\centering
\begin{tabular}{ c | c | c }
 Variable & Variable name & Range \\ 
 \hline
 x1 & Wall Insulation Thickness & 0m - 0.5m \\  
 x2 & Roof Insulation Thickness & 0m - 0.5m \\  
 x3 & Ground Insulation Thickness & 0m - 0.1m \\  
 x4 & Window Size & 20\% - 100\% \\  
 x5 & Window Overhang Length & 0\% - 100\% \\  
 x6 & Window Opening Amount & 0\% - 100\% \\  
 x7 & Roof Emissivity & 0.4 - 1 \\  
 x8 & Glazing Type & Double or Triple \\      
\end{tabular}
\caption{Table showing the inputs and their considered ranges for the building model.}
\label{tab:inputs}
\end{table}

Initially, in the first wave, we fit the two emulators using an initial set of 500 simulations: 250 unique $\textbf{x}$ input points chosen by a sliced Latin hypercube design \citep{Ba2015}, each replicated once. We then calculate $I_{oh}(\textbf{x})$ and $I_{eu}(\textbf{x})$ (that is, the implausibilities for the overheating risk and the energy usage) for each of the initialised candidate buildings.

As a comment, we are interested in the values of the overheating \emph{risk} and the \emph{average} energy usage, not the raw outputs of the simulator. Therefore, in calculating $I_{oh}(\textbf{x})$ via Equation \ref{eq:Impl}, $y(\textbf{x})$ is replaced with the logit overheating risk and $y(\textbf{x})$ is replaced with the mean energy usage in the calculation of $I_{eu}(\textbf{x})$. Because the logit overheating risk is used, rather than the overheating risk itself, we also modify the value of $L$, the target threshold, to be $\mathrm{logit}(0.01)$ rather than just 0.01. The logit overheating risk is the original output of the latent Gaussian process emulator, and is also unbounded. Using the overheating risk itself could also be done, although such quantity is bounded between 0 and 1.


With these two sets of implausibilities, $I_{oh}(\textbf{x})$ and $I_{eu}(\textbf{x})$, we take the overall implausibility as $I(\textbf{x}) = max(I_{oh}(\textbf{x}), I_{eu}(\textbf{x}))$. Any candidate building $\textbf{x}$ where the overall implausibility is greater than 3 can then be ruled-out (and any less than $-3$ can be ruled-in). 

As discussed previously, this process can then be repeated. For the next wave's simulation data, a random selection of 250 buildings are chosen from the larger candidate set, only considering those which are still NROY/NRIY. Each of these chosen buildings is then simulated twice. This simulated data set, along with any of the previously simulated data which is still NROY, can be used to re-fit the emulators; providing improved accuracy for buildings believed to be near the boundary. These newer emulators can then be used to further rule-out (or rule-in) candidate buildings. A key computational attribute here is that once a building is ruled out (or ruled in), it no longer needs to be checked - it has already been ruled out (or ruled-in), its final implausibility value is the last one it was assigned. 

We performed three waves of this history matching inspired level-set estimation. In wave 1, $14.47\%$ of the space was found with $I(\textbf{x}) < 3$ (i.e. $85.53\%$ of all the initial candidate buildings were immediately ruled-out / only $14.47\%$ were left NROY) and $0.10\%$ of the space was found with $I(\textbf{x}) < 0$ (i.e. a very small $0.10\%$ of the initial candidate buildings were found to be tenably future-proof). By wave 2, the NROY space was shrunk further down to $7.44\%$ of the total space and $1.49\%$ of the space was tenable. By the third wave, $6.32\%$ was NROY and $1.99\%$ of all buildings were tenably future-proof. To summarise, wave 1 was essentially unable to find any viable building designs, but it was able to clearly rule-out many designs. Later waves were able to leverage this information by simulating more densely in the not ruled-out yet space, allowing many viable building designs to be discovered.

To visualise the types of buildings which are most future-proof, we make use of standard (in the history matching literature) minimum implausibility and optical depth plots \citep{Andrianakis2017}. For every combination of two input variables, a 2D grid is made. Every candidate building is then sorted into the relevant grid cell for the 2D combination of input variables. Minimum implausibility plots present the minimum implausibility of any building within each grid cell, and the optical depth plots plot the proportion of buildings that are NROY within each grid cell. Small minimum implausibilities or large optical depths represent good building design choices. These plots then provide information about the shape of the NROY space and the implausibility in this 2D projection. This can then be repeated for all 2D input variable combinations. These plots provide insight into what types of buildings are acceptable, and key relationships can be identified. We also present the 1D histograms showing the proportion of buildings which are NROY for each input. These make it more clear what the individual impact of each input is, but it is important to remember that the overall shape of NROY is a complicated 8D surface with potentially many complex interactions.

\begin{figure}[!htb]
\includegraphics[width=\textwidth]{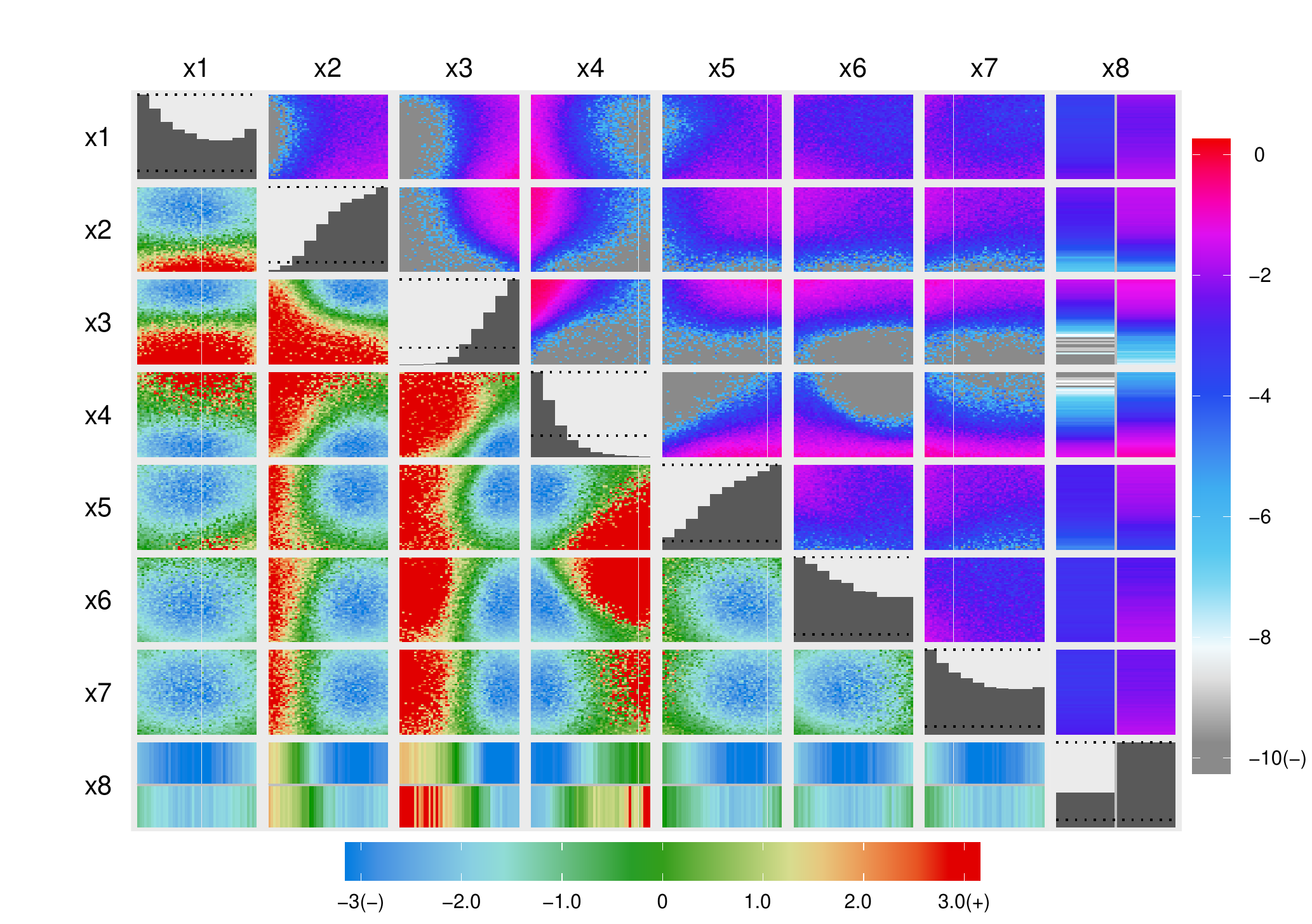}
\caption{Minimum implasubility plots (below and left of diagonal) and optical depth plots (above and right of diagonal) for wave 3. For the minimum implausibility, the scale is capped above by 3 (above this all buildings are ruled-out), and below by -3 (below this all buildings are ruled-in). For the optical depth, the log scale is used, and goes from 0 (i.e. all buildings are non-implausible) down to -10 or lower (i.e. less than $0.0045\%$ of buildings are non-implausible). The diagonal histograms show the \emph{relative} proportion of buildings which are NROY. Also shown are two horizontal lines indicating how much the histograms should be scaled down to instead present the \emph{absolute} proportion of buildings which are NROY.}
\label{fig:Wave3}
\end{figure}

Figure \ref{fig:Wave3} presents the results from wave 3.

From the figure, we can see that small values of $x_2$, $x_3$ and $x_5$, large values of $x_4$, and $x_8$ equal to 0, are all clearly poor choices for this building. To a lesser extent large values of $x_1$, $x_6$, and $x_7$ are also poor choices. There are also several key 2D interactions that can be observed - one example is that between $x_4$ and $x_5$, where a larger value of $x_4$ can be acceptable if there is also a larger value of $x_5$. Other interesting patterns are also visible, for example, between $x_2$ and $x_3$, or $x_4$ and $x_6$.

From this point, we leave a final choice of building design down to a practitioner because secondary (or in this case, tertiary) criteria often exist. Preferably, the practitioner would choose only from ruled-in designs, but this can be an overly strict criteria and barely any ruled-in buildings have been found ($0.0019\%$ of buildings have $I(\textbf{x}) < -3)$ by the end of the third wave). Additional waves with further simulation would rectify this, but such a strict requirement is not always necessary. If a large number of additional waves were performed, it is possible that the number of NROY/NRIY buildings would approach 0; with all candidate buildings either ruled-out or ruled-in. In these scenarios, several methods exist for sampling new points from the NROY (/NRIY) space (such as the method from \cite{drovandi2017new}), allowing the level-set boundary to be more precisely defined. We advise against this however, as there is a limit to how precisely a building can actually be built in practice, and so there does exist an upper limit to how many waves could be performed.

A bare-minimum requirement for selecting a final building might involve considering any buildings with $I(\textbf{x}) < 0$. These are all the buildings where the emulator mean predictions (i.e. the best guesses) suggest both requirements are satisfied. Using this requirement, and selecting the allowed design with the largest windows, leads to a final building which could be built. For this building, the wall insulation is 0.375m thick, the roof insulation is 0.458m thick, the ground insulation is 0.099m thick, the windows take up $94.2\%$ of their allowed maximum space, the overhangs are $98.1\%$ as long as the window heights, $12.4\%$ of the window area can be opened, the roof has an emissivity value of 0.672, and the windows are triple-glazed. However, on reflection, this building only has an implausibility value of $-0.057$ which is not very negative, indicating a lack of confidence that the building is in fact satisfactory. With the emulators, we can also calculate that this building has a $54.8\%$ chance of meeting the overheating criteria and a $56.4\%$ chance of meeting the energy usage criteria. Neither are particularly large. Additionally, since we have 2 criteria (treated independently), the joint probability of meeting both criteria is thus only $30.9\%$.

A stricter constraint, but not so strict as $I(\textbf{x}) < -3$, is thus a more sensible final requirement. Only considering buildings where $I(\textbf{x}) < -1$ (and choosing the building with the largest windows) leads to a building design with an implausibility of $-1.25$, a $93.2\%$ probability of meeting the overheating criteria, a $89.4\%$ probability of meeting the energy usage criteria, and an acceptable $83.3\%$ probability of meeting both criteria. This building has 0.293m of wall insulation, 0.402m of roof insulation, 0.096m of ground insulation, the windows take up $63.8\%$ of their allowed maximum area, the overhangs are $79.2\%$ of the window height, $6.9\%$ of the window areas are openable, the roof emissivity is 0.779, and the windows are triple-glazed.

Several possible criteria for deciding the final building design can be imagined. In practice, the decision depends on the priorities and desires of the practitioners involved, but the illustration above serves as a good example. The emulators built (and improved with the level-set methodology) provide the information needed to make such decisions, and provide insight into the various trade-offs a practitioner will have to balance.

\section{Conclusion}
\label{sec:conc}

To conclude, we presented a case study - using ideas from the wider Uncertainty Quantification community to improve upon a building design. After just three waves of simulation, $92.56\%$ of possible modifications were discarded as implausible. The procedure in general is accessible and allows an engineer to choose from a set of acceptable building designs according to other more subjective (or less tangible) requirements. Importantly, we also treated EnergyPlus as stochastic, which acknowledges more of the uncertainties present and could contribute to shrinking the observed performance gap between simulated buildings and constructed buildings \citep{imam2017building}.

Throughout, we have ignored the notion of `model discrepancy', where one acknowledges the simulator is not perfect and is itself flawed. It is straightforward and common to add an additive, constant, zero-mean measure of this discrepancy in history matching \citep{Vernon2014, Andrianakis2015}, by simply replacing the variance term in the implausibility equation, $V(y(\textbf{x}))$, with $V(y(\textbf{x})) + V_{MD}$, where $V_{MD}$ represents the subjective uncertainty around what the difference between the simulated quantity and the real world quantity could be. If $V_{MD}$ is not believed constant, or additive, or zero-mean (all possible within a level-set estimation procedure), then more must be done. Model discrepancy is an important question when it comes to any form of simulator analysis \citep{Goldstein2009}. Whether (and how) to include model discrepancy into building design requirements is a question that requires discussion between statisticians, practitioners and policy makers.

Additionally, the level-set methodology we use is certainly not the only one available. The problem of efficient simulator level-set estimation is an open research question, and other efforts exist \citep[see][for a review of some alternatives]{Lyu2018}. A comprehensive comparison between history matching techniques in general and alternatives is sorely missing from the literature, and remains an important topic for future research. In practice, history matching techniques serve as batch design schemes which are easily implemented and understood for a wide range of problems, whilst also being fairly robust due to the ruling-out process and the conservative nature. For our problem, the technique proves to be capable and can provide value to practitioners.

This article makes reference throughout to the ease and intuitiveness of history matching, and the history matching inspired level-set estimation methodology. This does not however mean that constructing emulators is always easy. Constructing an emulator requires careful assessments of potential assumptions. Any subsequent analysis (be that level-set estimation, prediction, optimisation, calibration, etc.), depends on this careful emulator construction, lest any conclusions be invalid. Gaussian processes are fairly complex (although implementation software for them is widespread); obtaining flexible probabilistic models of the quantities of interest. Simpler methods could instead be used, although this is likely to come with a decrease in accuracy or efficiency (\cite{salter2016comparison} tried this for early standard history matching waves and found the strategy to be detrimental).

For the two stochastic outputs we considered (energy usage and overheating classification), we were interested in improving specific summary statistics for each; the mean energy usage and the overheating probability. For the binary overheating classification output, the overheating probability fully summarises the output. For the energy usage however, summaries other than the mean could have been used with a different interpretation. In this case, because buildings are often used for many years and the energy usage for any one individual year isn't particularly important, the mean average is a sensible quantity of interest. For many problems however, other quantities of interest can be more important, such as the median or the $90\%$ quantile. The outlined level-set procedure can still work in these situations, although different emulators might be more suitable, such as a quantile Kriging emulator \citep{plumlee2014building}.

Overall, we do believe that emulation, the described level-set methodology, and the ideas discussed more generally, can be useful tools for the field of building performance simulation. Practical improvements to existing or planned buildings could easily be facilitated by utilising the tools outlined here.

\bibliographystyle{apalike}
\bibliography{references}

\appendix

\section{Appendix - Earlier Waves}

\begin{figure}[!ht]
\includegraphics[width=\textwidth]{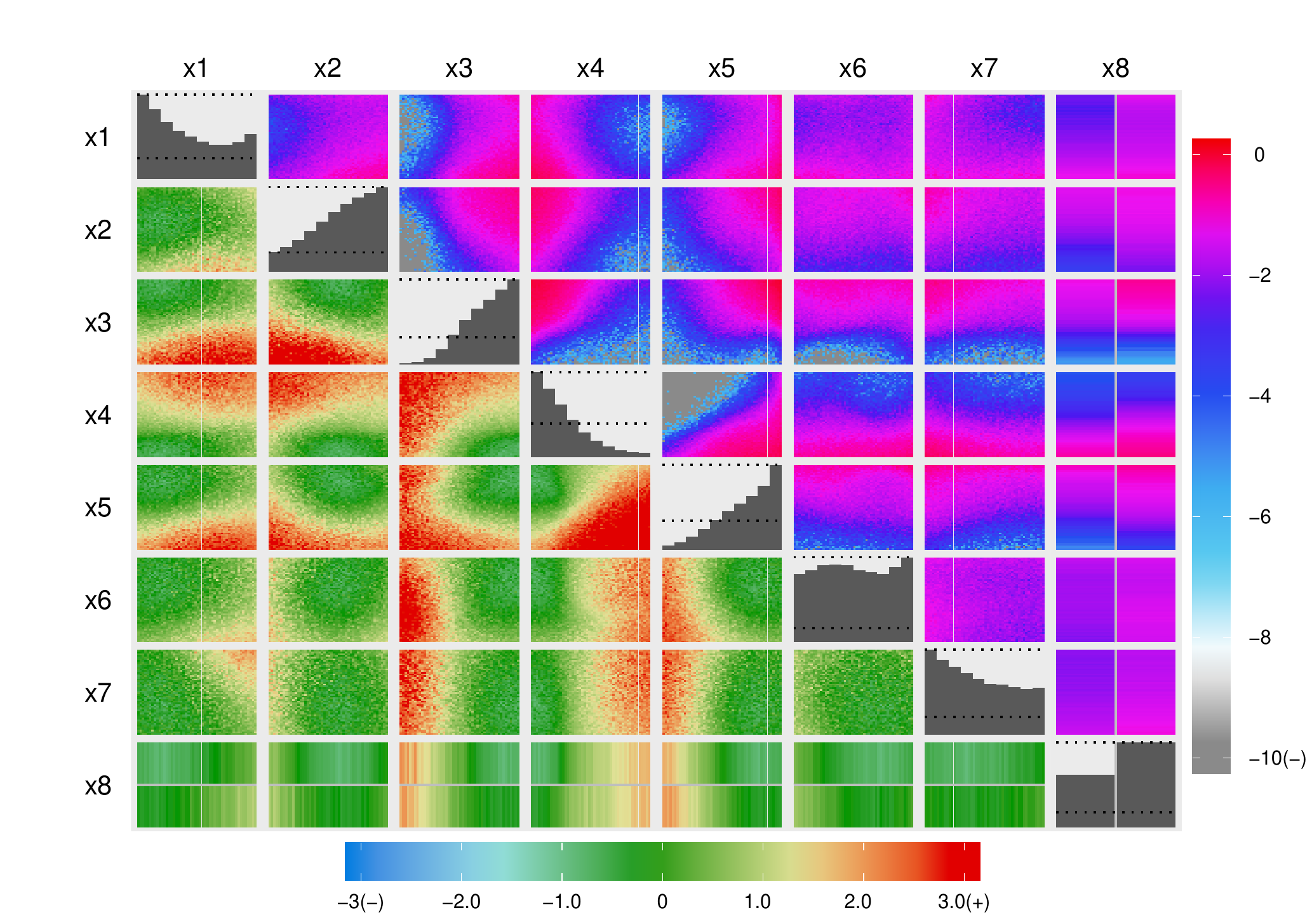}
\caption{Minimum implasubility plots (below and left of diagonal) and optical depth plots (above and right of diagonal) for wave 1. For the minimum implausibility, the scale is capped above by 3 (above this all buildings are ruled-out), and below by -3 (below this all buildings are ruled-in). For the optical depth, the log scale is used, and goes from 0 (i.e. all buildings are non-implausible) down to -10 or lower (i.e. less than $0.0045\%$ of buildings are non-implausible). The diagonal histograms show the \emph{relative} proportion of buildings which are NROY; also shown are two horizontal lines indicating how much the histograms must be scaled down to instead present the \emph{absolute} proportion of buildings which are NROY.}
\label{fig:Wave1}
\end{figure}

\begin{figure}[!ht]
\includegraphics[width=\textwidth]{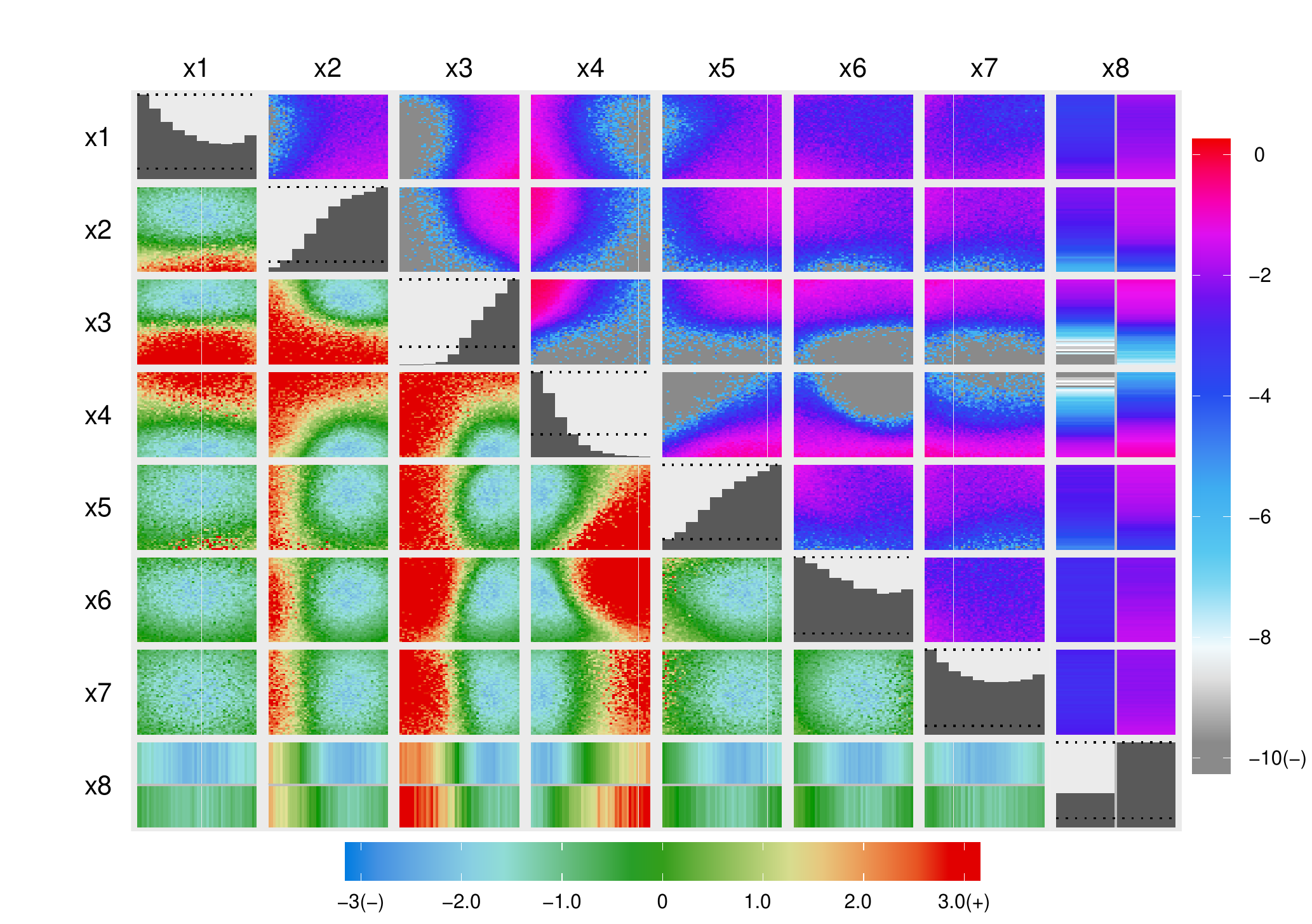}
\caption{Minimum implasubility plots (below and left of diagonal) and optical depth plots (above and right of diagonal) for wave 2. For the minimum implausibility, the scale is capped above by 3 (above this all buildings are ruled-out), and below by -3 (below this all buildings are ruled-in). For the optical depth, the log scale is used, and goes from 0 (i.e. all buildings are non-implausible) down to -10 or lower (i.e. less than $0.0045\%$ of buildings are non-implausible). The diagonal histograms show the \emph{relative} proportion of buildings which are NROY; also shown are two horizontal lines indicating how much the histograms must be scaled down to instead present the \emph{absolute} proportion of buildings which are NROY.}
\label{fig:Wave2}
\end{figure}

Section \ref{sec:results} did not present the intermediate wave implausibility plots, and so they are presented here. Figure \ref{fig:Wave1} presents the results from wave 1 and Figure \ref{fig:Wave2} presents the results from wave 2. In the earlier waves, fewer pixels show negative minimum implausibility and fewer buildings are ruled-out. The general shapes and patterns are still apparent in all waves, but visible changes are also made between the waves. 

\section{Appendix - Overheating Criteria}
In Section \ref{sec:simul} we discussed how we use an overheating metric defined by CIBSE \citep{cibse201352}, which we will outline here. Any different metric could equally have been used. The weighted running mean $T_{rm}$ for every day is first calculated as $T_{rm} = 0.2T_{od-1} + 0.8T_{rm-1}$, where $T_{od-1}$ is the outdoor temperature from the previous day, and $T_{rm-1}$ is the weighted running mean from the previous day. The maximum acceptable indoor temperature $T_{max}$ is then defined as $0.33T_{rm} +21.8$. This maximum acceptable temperature is then used to asses the relative temperature inside the building; $\Delta T$ is an hourly timeseries of the difference between the operative indoor temperature $T_{op}$ and the maximum comfort temperature $T_{max}$. $\Delta T$ then represents a value for how hot the building is at any one hour. $\Delta T$ can be calculated from the outputs from EnergyPlus, as EnergyPlus outputs the operative temperature and the outdoor temperature is a required input.

Three criteria are then constructed from $\Delta T$ to quantify the different ways a building can overheat. Criteria 1 is broken if more than $3\%$ of occupied hours between May and September have a $\Delta T$ value over 1 degree. This criteria checks how often the building is too hot.

Criteria 2 is broken if, for any given day, $\Delta T$ sums to be greater than or equal to 6 (only counting occupied hours). This criteria is a combination of the length of time a building is too hot, and how hot it is.

Criteria 3 is broken if, for any occupied hour, $\Delta T$ is greater than 4. This criteria checks how hot a building gets at its peak.

A building is then considered to overheat if 2 or more of the criteria are broken.

\section{Appendix - Validation}

History matching is based on the notion that regions of simulator space which obviously do not contain the answer need not receive attention. This is what motivates the conservative threshold used to rule-out (and in our case rule-in) input settings, and provides some degree of robustness to emulator error. Nonetheless, it is important that the emulators used still have a reasonable fit to the simulator data.

To check each wave's emulators, we obtained 400 new out-of-sample validation simulations (chosen by a sliced Latin hypercube). We then use these to check the emulators' predictive performance.
For the energy usage emulators, we obtain the 2 standard deviation intervals, and count the percentage of validation points that lie within. Close to $95.7\%$ of valdiation points should be contained within the intervals. For the first wave emulator 367 points lie within (or $91.75\%$); for the second wave 374 points lie within ($93.35\%$); and for the third wave 392 points lie within ($98.00\%$). All of these seem acceptable (perhaps slightly overconfident in earlier eaves and underconfident in later waves), and provide some credibility to the energy usage emulators.

For the overheating emulators, which are for binary outputs, we obtain the ranked probability scores \citep{epstein1969scoring, hersbach2000decomposition}. These are scores for how accurate probabilistic forecasts are. To then check if these observed ranked probability scores are adequate, we obtain reference distributions; sampling multiple hypothetical data sets from the underlying emulators and using these to score our emulators. Then, if the observed ranked probability scores do not seem out of place when compared to the reference distributions, then confidence is gained in the overheating emulators.
For the first wave emulator we obtain a ranked probability score of 0.159, which is less than the $95\%$ sample quantile from the reference distribution (0.174); the second wave emulator provides a ranked probability score of 0.029 which is also less than the $95\%$ reference quantile (0.040); and the third wave emulator provides a ranked probability score of 0.020 which is less than the $95\%$ reference quantile (0.025). This provides evidence that the overheating emulators are acceptable, and the emulators collectively are acceptable.

\end{document}